# Optimized Data Loading for a Multi-Terabyte Sky Survey Repository*

Y. Dora Cai[†], Ruth Aydt[†], Robert J. Brunner[†,‡]

[†]National Center for Supercomputing Applications (NCSA)
[‡]Department of Astronomy
University of Illinois at Urbana-Champaign
{ycai, aydt, rb}@ncsa.uiuc.edu

## Abstract

Advanced instruments in a variety of scientific domains are collecting massive amounts of data that must be post-processed and organized to support scientific research activities. Astronomers have been pioneers in the use of databases to host highly structured repositories of sky survey data [13]. As more powerful telescopes come online, the increased volume and complexity of the data collected poses enormous challenges to state-of-the-art database systems and data-loading techniques. When the data source is an instrument taking ongoing samples, the database loading must, at a minimum, keep up with the data-acquisition rate. These challenges are being faced not only by the astronomy community, but also by other scientific disciplines interested in building scalable databases to house multi-terabyte archives of complex structured data.

In this paper we present *SkyLoader*, our novel framework for fast and scalable data loading that is being used to populate a multi-table, multi-terabyte database repository for the Palomar-Quest sky survey [8]. Our framework consists of an efficient algorithm for bulk loading, an effective data structure to support data integrity and proper error handling during the loading process, support for optimized parallelism that matches the number of concurrent loaders with the database host capabilities, and guidelines for database and system tuning. Performance studies showing the positive effects of the adopted strategies are also presented.

Our parallel bulk loading with array buffering technique has made fast population of a multi-terabyte repository a reality, reducing the loading time for a 40-gigabyte data set from more than 20 hours to less than 3 hours. We believe our framework offers a promising approach for loading other large and complex scientific databases.

*This work was supported in part by the National Science Foundation grants SCI 0525308, ACI-9619019, ACI-0332116 and by NASA grants NAG5-12578 and NAG5-12580.

## 1. Introduction

With the advent of computers, databases, data warehouses, and World Wide Web technologies, astronomy research has been undergoing revolutionary changes. Advanced data-gathering technologies have collected tremendous amounts of digital sky survey data and many sky survey repositories (e.g., SDSS, GALEX, 2MASS, GSC-2, DPOSS, ROSAT, FIRST and DENIS) have been built to house this data and to serve as valuable resources for astronomy researchers and the general public [13]. Palomar-Quest is one of the sky survey repositories currently under construction.

Several characteristics of sky survey data—such as rapid data capture, massive data volume, and high data dimensionality—make data loading the first great challenge in building advanced sky survey repositories. These characteristics are reflected in several demanding issues that must be addressed when loading such data. First, data-loading speed must keep up with data-acquisition speed. Second, it must be possible to populate multiple database tables from a single source file. Third, it is often necessary to perform complex data transformations and computations during the loading process. Finally, automatic error recovery is required during the lengthy data-loading process.

In this paper we present *SkyLoader*, our optimized framework for parallel bulk loading of a Palomar-Quest repository powered by an Oracle10g relational database. Our framework addresses all of the data-loading issues listed above through the development and application of the following techniques: (1) an efficient algorithm to perform bulk data loading, (2) an effective data structure to maintain table relationships and allow proper error handling, (3) optimized parallelism to take full advantage of concurrent loading processes, and (4) active database and system tuning to achieve optimal data-loading performance.

With systematic testing and refinement of the *SkyLoader* framework we have significantly improved the data-loading performance for the Palomar-Quest repository. Loading time for a 40-gigabyte data set was reduced from more than 20 hours to less than 3 hours on the same hardware and operating system platform. We firmly believe the experience gained in this study will benefit other data repositories of massive scale.

The remainder of the paper is organized as follows. Section 2 presents a brief introduction to the Palomar-Quest sky survey. Section 3 describes data-loading challenges and approaches. Section 4 details our *SkyLoader* framework for addressing these data-loading challenges. Section 5 describes our experimental platform and presents results and analyses for a variety of performance studies. Section 6 discusses our approach in comparison to related work, and Section 7 presents conclusions and future directions.

## 2. The Palomar-Quest Sky Survey

The Palomar-Quest sky survey is a collaborative endeavor between the California Institute of Technology, Yale University, Indiana University and the University of Illinois, being jointly led by Principal Investigators Charles Baltay at Yale and S. George Djorgovski at Caltech. Palomar-Quest is a multi-year, multi-wavelength synoptic survey conducted at the Palomar-Quest Observatory located in north San Diego County, California. The survey camera consists of 112 Charge-Coupled Devices (CCDs) and can observe a third of the sky in a single night. In contrast to traditional sky surveys, Palomar-Quest repeatedly scans the night sky. If we characterize traditional sky surveys as taking digital snapshots of the sky, the Palomar-Quest survey in contrast is taking digital movies [8]. The time element inherent in this survey allows astronomers to statistically analyze the variable nature of our universe and contributes to the volume, richness, and complexity of the survey data.

The data-collection rate is 7.4 gigabytes/hour or approximately 70 gigabytes/night, with a monthly average of 12–15 nights of observing. Extrapolating, Palomar-Quest can collect approximately 1 terabyte of image data per month, assuming ideal observing conditions. The image data captured by the telescope camera is further processed to produce catalog data totaling approximately 15 gigabytes/night. Since going into production in 2003, over 6 terabytes of raw image data have been archived at NCSA from which more than a terabyte of catalog data has been derived.

Researchers from the Department of Astronomy and the National Center for Supercomputing Applications at the University of Illinois at Urbana-Champaign have jointly designed and developed a data repository system powered by an Oracle 10g relational database to archive, process, and distribute the Palomar-Quest sky survey catalog data to research collaborators. *This paper focuses on the optimized loading of derived catalog data into the sky survey repository.*

## 3. Data Loading Challenges and Approaches

The large data-collection rates and volumes noted in the previous section dictate the necessity for a fast data repository loading process that is capable of keeping up over time with the speed of data acquisition. A number of factors contribute to the difficulty of achieving this goal.

Collected raw image data and computed catalog data are usually archived in a mass storage system that is separate from the database server. The catalog data that must be transferred from the mass storage system to load the database repository typically saturates the available network bandwidth, introducing the network as the first bottleneck to fast data loading.

Sky survey data encompasses information of many different types, from sky region specifications to the observed details of tiny objects. This variety of information is interleaved in the catalog data set that is generated when the raw image data is processed. During the data-loading process the complex catalog data must be parsed, the correct destination tables must be identified, and the data must be loaded into multiple target tables in the repository. Loading data into multiple tables is further complicated by the presence of multiple relationships among tables—relationships that must be maintained by complying with the primary and foreign key constraints during the loading process.

Additional operations are also performed during the data-loading process. These operations include transformations to convert data types and change precision, validation to filter out errors and outliers, and calculation of values such as the Hierarchical Triangular Mesh ID (htmid) and sky coordinates to facilitate the science research [10] that the repository is built to enable. All such intensive operations place an additional burden on the loading process. Finally, since data loading is typically a lengthy process, a mechanism of automatic recovery from errors is a basic requirement.

Each major relational database management system (RDBMS) vendor provides a utility to load data from a flat file into a table. The Oracle system supports SQL*Loader, the MS/SQLServer system supports Data Transformation Services (DTS), and IBM DB2 supports a LOAD utility. However, these tools are proprietary tools that can only work with the vendor's own databases. Furthermore, they are primarily designed to quickly load data into a single database table without performing any data transformation. These data-loading services are not suitable for use with massive scale sky survey data.

Several packaged data-loading tools are available on the market, such as BMC Fast Import, FACT (CoSORT's FAst extraCT), and DataSift. However, these data-loading tools are black boxes that generate programs which cannot be easily customized [1]. Some new bulk-loading techniques have been proposed [1, 4, 5, 9, 11]; however, all of these approaches are focused on bulk loading an index, such as B++-tree, Quad-tree and R-tree. Based on our experience and examination of the research literature, there is little work on parallel bulk loading of huge amounts of data into a multi-table database.

To meet the challenges in building the Palomar-Quest repository, we have designed and implemented an optimized framework, called *SkyLoader*, which consists of (1) an efficient algorithm to load data in bulk, (2) an effective data structure to maintain table relationships and handle errors, (3) optimized parallelism to take full advantage of concurrent loading processes, and (4) active database and system tuning to achieve optimal data-loading performance. Using this framework we can bulk load data in parallel, insert data into multiple database tables simultaneously without locking and constraints violations, and recover the loading process from errors. The *SkyLoader* framework has significantly improved the performance of data loading. We have been able to reduce the loading time for a 40-gigabyte data set from over 20 hours to less than 3 hours on the same hardware and operating system platform.

## 4. The SkyLoader Framework

In this section we present the design of our sky survey repository and the details of our *SkyLoader* framework.

### 4.1 Data Model and *SkyLoader* Tasks

The raw images captured by the camera on the Palomar-Quest telescope are archived in a UniTree mass storage system. A program is run on the raw image data to extract catalog data, which includes a wide range of information. Typically the catalog data includes information on the telescope position, the sky region scanned, the parameters applied, the CCDs operated, the frames derived, and the objects captured. The catalog information is first written to an ASCII file, which is saved in the mass storage system and then uploaded to a repository database. The format of catalog file varies depending on the extraction program used. In most cases, different aspects of the catalog information are interleaved in the file. For example, a row of *frame* information is followed by four rows of *frame aperture* information, and a row of *object* information is followed by four rows of *finger* information. Usually each row in the catalog data file has a tag or a keyword that can be used to determine the destination table in the database.

A commercial relational database, Oracle 10g, has been chosen to host the Palomar-Quest repository. The repository database has been designed to store the catalog data and support data analysis. Figure 1 shows the data model for the database, which consists of 23 tables. Only the table names and relationships are shown in Figure 1 to simplify the diagram while still conveying the complexity of the model and inter-table relationships.

Each table stores a unique aspect of the sky survey. For example, metadata related to a night's observation such as telescope position, filters in use, and collection start time goes into the *observations* table. Metadata related to the CCDs such as CCD number and sky area covered goes into the table *ccd_columns*. Detailed information related to observed objects goes into the *objects* table.

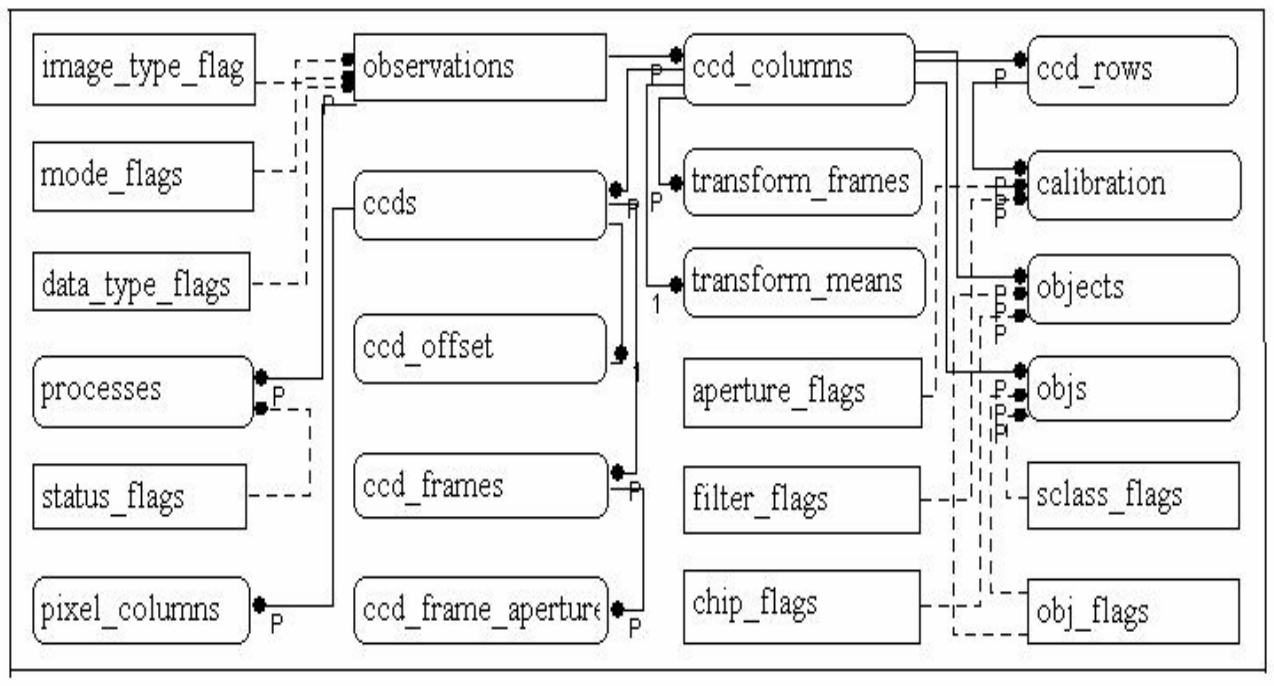

**Figure 1. Palomar-Quest Repository Data Model**

A primary key is defined in each table to force data uniqueness. Most tables have one or more foreign keys to maintain parent-child relationships. For example, a frame aperture is always related to a frame. The foreign key on the table *ccd_frame_apertures*, which references the table *ccd_frames*, enforces this constraint. The database table sizes vary significantly. Some static metadata tables have less than 100 rows, while the *objects* table is expected to grow beyond a billion rows.

Taking into account the data model and data characteristics of the Palomar-Quest repository, the *SkyLoader* framework was designed to efficiently perform the following tasks using a parallel architecture: (1) read the data from the catalog data files, (2) parse, validate, transform and compute data, (3) load data into the repository database and distribute data to multiple tables, and (4) detect and recover from errors in the data-loading process.

### 4.2 An Efficient and Scalable Bulk-Loading Algorithm

For massive volumes of sky survey data, it is crucial to explore scalable data-loading techniques. The first such technique to explore is *bulk loading*. Bulk loading allows multiple *insert* operations to be packed into a single batch and performed with one database call, minimizing network roundtrip traffic and disk I/O [17].

It is straightforward to perform bulk loading to a single table, and most RDBMS system tools and some on-the-shelf software packages can accomplish this efficiently. However, it is nontrivial to bulk load multiple tables simultaneously due to the complicated relationships among the tables. If the data belonging to a child table is loaded before the corresponding parent keys, a foreign key constraint is violated. Our technique to avoid this problem is to first buffer the data into separate arrays designated for different tables, and then to follow the parent-child relationship order when performing the bulk inserts. The parent table is loaded first, then the child table(s). This table-loading order is illustrated in Figure 2.

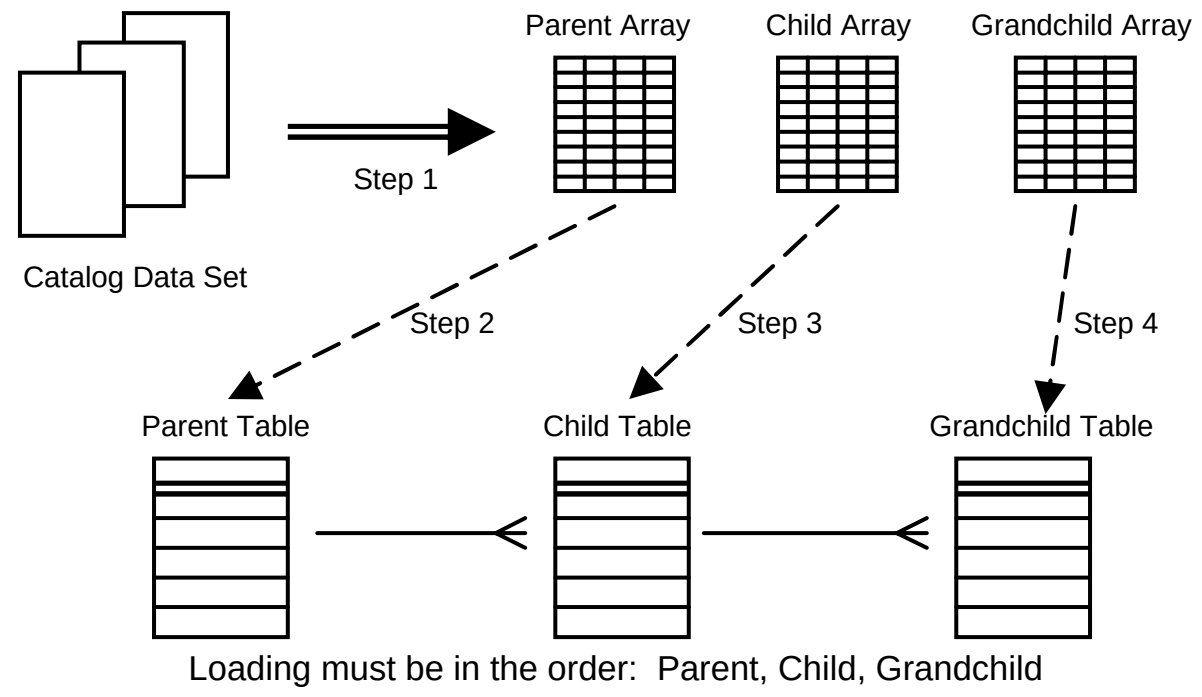

**Figure 2. Bulk Loading Order with Multiple Tables**

Another difficulty in data loading is recoverability in a lengthy data-loading process. The catalog data set to be loaded sometimes contains errors such as missing and/or invalid values. To make the loading process recoverable from these errors, we use an array-index tracing technique that can quickly detect errors, skip the problematic rows, and resume the loading process immediately.

For our *SkyLoader* framework we developed an efficient algorithm, *bulk-loading*, that enables bulk loading into multiple tables. This algorithm not only speeds up data loading by a factor of 7 to 9, but also maintains the relationships of multiple tables and enables the system to recover from errors during data loading. Our *bulk-loading* algorithm is presented in Figure 3.

The algorithm, *bulk-loading*, contains two user-tunable constants, *array-size* and *batch-size*, controlling the size of an array and the size of a batch, respectively. The procedure *bulk_loading* (Line 4, in Figure 3) first parses a data row, performs validation, transformation, and computation, and then buffers the data row into a designated array. This buffering step separates data into different arrays based on the destination tables and is necessary to maintain the relationships between multiple tables and to facilitate error handling. We explain this step in more detail in the next section on our buffering data structure.

When any data array reaches *array-size* (Line 5), the *batch_row* procedure is called (Line 10) for each array based on the parent-child relationship. The array for the parent table is processed first, followed by the child tables. This processing sequence depends entirely on the data model.

```
Input: a series of input data files
Output: populated database tables

int array-size   /* the size of an array */
int batch-size   /* the size of a batch;
                    typically << array-size */

Procedure bulk_loading
(1) for each data file {
(2)   open the file
(3)   for each row {
(4)     parse the row, do validation, transformation and computation, and buffer it in a designated array based on the destination table;
(5)     if (any array.size >= array-size) {
(6)       for each array ordered by parent-child relationship {
(7)         first_idx = 0;
(8)         last_idx = array.size;
(9)         while (first_idx <= last_idx)
(10)          first_idx = batch_row(array, destination_table, first_idx, last_idx)
(11)      } /* for each array */
(12)    } /* if reach array-size */
(13)  } /* for each row */
(14) } /* for each data file */

Function batch_row (array, destination_table, first_idx, last_idx)
(15)  while (first_idx <= last_idx) {
(16)    prepare SQL statement;
(17)    add to batch;
(18)    if (batch-size reached) {  /* time to insert */
(19)      insert batch into the destination table;
(20)      if (successful insert) {
(21)        first_idx += batch-size;
(22)      } else {      /* if an error occurred skip that row */
(23)        skip_one_row;
(24)        return (the_next_index);
(25)      }
(26)    } else if ( first_idx == last_idx) {  /* array done */
(27)      insert batch into the destination table;
(28)      if (successful insert) {
(29)        return (last_idx + 1 );
(30)      } else {      /* if an error occurred skip that row */
(31)        skip_one_row;
(32)        return (the_next_index);
(33)      }
(34)    }
(35)  } /* while there are more rows to process */
```

**Figure 3. *bulk-loading* Algorithm**

The function *batch_row* prepares the SQL statements (Line 16), adds the SQL command to a batch (Line 17) and makes a database call when *batch-size* is reached (Line 19). If no error is encountered (Line 21), the function loops through the array and inserts all rows in batch into the appropriate database table (Line 27). In the case of errors, the function skips the error row, and returns the next array index to the calling procedure *bulk_loading* (Lines 23, 24 and 31, 32). The *bulk-loading* procedure continues to make calls to the function *batch_row* with a new index range (Line 9 to Line 10) until all rows in the array are processed.

**Example 1: Loading a data set into two tables.** Suppose a sky survey repository has two tables: *frames* and *objects*. The table *frames* stores the frame information derived from an image and the table *objects* keeps the object information measured in each frame. A foreign key constraint between *frames* and *objects* is enforced. Consider a case with 5 frames and 1000 objects interleaved in a data file. When packaging the SQL insert statements into a batch for bulk loading, the rows with *object* information will reach *batch-size* first. However, if the *object* rows get inserted before the *frame* rows, the *frames-objects* foreign key constraint will be violated immediately. Using our *bulk-loading* algorithm, a *batch-size* of 40, and an *array-size* of 1000, this data set can be loaded efficiently and correctly through the following steps:

**Step 1**. Read data in and buffer the *frame* data into *array1* and the *object* data into *array2*.

**Step 2**. When either array reaches *array-size*, 1000 in this example, bulk loading is triggered. In this example, *array2* will fill up first. Despite that, the bulk loading proceeds by following the parent-child relationship order, meaning the rows in *array1* are processed before *array2*.

**Step 3**. If no error occurs, a single call to the function *batch_row* will initiate bulk loads to insert all rows in an array into a database table. Suppose row 45 in *array2* has an error. Recall, we are using a *batch-size* of 40. The function *batch_row* inserts rows 1 to 40 in the first batch, inserts rows 41 to 44 in the second batch, skips row 45 where the error occurs, and returns to the calling procedure, *bulk_loading*. Since the array was not completely loaded, *bulk_loading* calls *batch_row* again for *array2* with a new starting index, and loading proceeds with rows 46 to 85, rows 86 to 125, and so on until all remaining rows in *array2* have been inserted.

The *bulk-loading* algorithm has been implemented using the JDBC™ core API. Let *N* denote the total number of rows in the data set. In the best case, that is when the data set is error-free, the algorithm will generate *N*/*batch-size* database calls and result in *N*/*batch-size* database I/Os. In the worst case, for example primary key violations on every row caused by repeatedly loading duplicate rows, bulk loading will deteriorate to a series of singleton *insert* operations which make *N* database calls and perform *N* database I/Os. This behavior results from the algorithm breaking up the problematic batch, skipping the error row, and repacking the batch to continue each time that an error is encountered.

Performance results demonstrating the benefits of bulk loading for our Palomar-Quest repository are shown in Section 5.1. The effects of the user-tunable constants *batch-size* and *array-size* are presented in Sections 5.2 and 5.3 respectively.

## 4.3 An Effective Data Structure to Buffer Data

As discussed previously, the catalog data set used to load the Palomar-Quest repository contains rows of data destined for multiple target tables in the database. This interleaving of data for multiple target tables, combined with the presence of multiple relationships among tables—relationships that must be maintained by complying with the primary and foreign key constraints during the loading process—makes bulk loading especially challenging. To manage the interleaved data and complex table relationships, and to facilitate quick recovery when an error is detected during the data-loading process, we have designed an effective data structure, *array-set*, in our framework.

The *array-set* data structure consists of a dynamically maintained set of two-dimensional arrays, each associated with a destination table in the database. One dimension of each array corresponds to table rows, and the other to table attributes. Arrays are cached in memory and used in the bulk loading process as described in the previous section.

The number of arrays in the *array-set* at a given time during data loading depends on the degree to which the data in the catalog data set is interleaved. As the input catalog data set is processed, the framework creates a new array in *array-set* whenever it reads an input row targeted for a database table for which no array is currently maintained. When any of the arrays in *array-set* are fully populated, bulk loading occurs. At the end of the bulk-loading cycle, the arrays in *array-set* are destroyed and their memory released. The framework resumes reading the input catalog data and creates new arrays as required to buffer the incoming table rows.

To reiterate and expand on the motivations for the *array-set* structure, the Palomar-Quest catalog data set contains various levels of information for a sky survey, and that information is interleaved in a single data set with the relationships between levels embedded in the file. If data is bulk-loaded by simply following the order of the data rows and starting bulk loads into various tables when a threshold is hit, a foreign key constraint may be violated because the referencing data may be loaded before the referenced data. In order to load the catalog data items into different destination tables and retain the proper relationships, we use *array-set* to buffer the data and execute the bulk loading in the order of parent-child sequence.

Error handling in bulk data loading is difficult. In the JDBC core API, when an error is encountered during a bulk load, the remaining data in the batch is ignored. Furthermore, after the batch has been dispatched to the database server, it is impossible to reapply it [12]. Since it is not unusual for sky survey data to have missing and/or invalid values due to the complexity of the collection and processing pipeline, stringent data checking is performed by the database to guard against hidden corruption, and errors are detected during bulk loads fairly often. Quickly recovering from an error in a single row and continuing to insert data in the batch following that row is critical. Failure to recover properly could result in the loss of a huge amount of information and in an incomplete catalog repository. Failure to recover quickly will negatively impact overall loading time.

The use of the *array-set* data structure allows us to solve these problems. Buffering data in an array enables random access of any data element. A row in a batch always maps back to the source array. By detecting the error row in a batch during bulk inserts, our algorithm can quickly identify the corresponding row in the source array, skip the error row, repack the batch, and continue the bulk data-loading process from the row following the error.

The tunable parameter *array-size* is one of the factors that effects bulk loading performance. A large *array-set* may consume too much memory on the client machine and cause excessive memory paging. This slowdown on the client where the loading process is running is reflected in degraded loading performance on the database server. On the other hand, an *array-size* value that is too small may increase the overhead for array initialization and population. In our framework, we adjusted *array-size* based on the system resources and data characteristics to achieve optimal performance. Results of the performance studies that we performed to select an optimal *array-size* value are presented in Section 5.3.

Our current *SkyLoader* framework uses a single *array-size* user-tunable constant to control the number of rows in all memory-resident arrays used to cache table data prior to bulk loading. Since the systems that we are using to run the client data-loading processes have generous memory configurations, our primary consideration was to implement a solution quickly rather than to carefully minimize the space needed by the *array-set* data structure. We plan to revisit this implementation and make use of a configuration file to support arrays with variable number of rows. By understanding the structure of our catalog data set and the interleave pattern of the rows there, we can make more intelligent memory-management decisions regarding the *array-set* data structure. We may also explore the use of an overall "memory high water mark" that would trigger bulk loading and memory reclamation whenever the aggregate memory used by the cached arrays reached that size.

## 4.4 Exploration of Optimized Parallelism

The Palomar-Quest survey collects a tremendous amount of data, with each observation generating 28 image data sets and each image data set containing the data collected by 4 CCDs. The raw image data is processed to derive the catalog data, which is also organized in 28 files. The 28 catalog data files can be processed independently, and we currently load them in parallel from Radium, an NCSA Condor [16] cluster, to a centralized Oracle database powered by an 8-processor SGI Altix server that is one of NCSA's TeraGrid resources. Parallelism enables multiple processors to work simultaneously with the database server and substantially improves the data-loading performance [2].

In our framework, we use two techniques to achieve the optimal degree of parallelism and minimize the wall-clock time required to load the data from an observation. First, we use as many Condor processes as possible to saturate the CPUs on the database server. Second, we recognize that the 28 catalog data files associated with an observation vary in size and, consequently, in loading time. Because of this variation, we assign unloaded data sets to the Condor nodes "on the fly" rather than dividing the data sets evenly among the Condor nodes. As soon a node completes the loading of one data file, another file is assigned to it until no unloaded catalog data files remain. This load-balancing methodology also helps minimize the overall data-loading time when one or more data files have a higher-than-average frequency of errors that slow the loading process.

The optimal degree of parallelism varies depending on the system resources and the running applications. In an ideal environment with our 8-processor database server and well-matched Condor nodes and network connectivity, we would expect 8 parallel loading processes to fully utilize all CPUs on the database server. However, our tests have shown that parallelism at this level tends to cause locking problems attributable to the fact that all RDBMS have a limit on the supported number of concurrent transactions. Even without the locking issues, the performance gain in data loading is usually not proportional to the degree of parallelism. Parallel processing introduces some system overhead that limits the performance benefit to less than perfect speedup.

In cases such as the Palomar-Quest sky survey repository where data loading is a critical and ongoing activity, it is worth the time to conduct careful experiments to determine the optimal degree of parallelism. Methodical experimentation—even when the detailed database system implementation is unknown—can help identify the best possible degree of parallelism. In our framework, we have parallelized the data loading according to the number of processors available, the underlying data characteristics, and the results of our performance studies. Performance study results are shown in Section 5.4.

## 4.5 Active Database and System Tuning

Many factors impact the performance of data loading. In our *SkyLoader* framework, we performed active database and system tuning to achieve the best possible configuration. Such performance tuning is crucial to achieve the fast loading of massive data volumes required for our sky survey repository. We believe our experience will benefit others who are faced with loading large quantities of scientific data from various disciplines into relational database systems.

### 4.5.1 Delay Index Building

Since the Palomar-Quest sky survey is a multi-year continuous effort, the survey repository must serve two purposes at the same time. First, it must be a warehouse to store incrementally loaded data. Second, it must act as a query engine to support scientific research. For optimal query performance, it is necessary to create indices on database tables. However, indices usually make data loading slower because every insert requires an update of all index entries [6].

Our tests showed that locking tends to happen more frequently at a lower degree of data-loading parallelism when indices are present. Based on our experiments, presented in Section 5.5, the impact of indices on data loading varies depending on the type and size of the index and on the pattern of the index keys. Because of these findings, we dropped most secondary indices to speed up the data loading. Once the catch-up phase of loading is complete and load time is not as critical, these secondary indices will be rebuilt and kept current as subsequent data is collected and loaded. Recognizing the need to balance load time and query performance, some very selective indices that are crucial to the scientific research queries, such as the index on htmid, have been maintained during the intensive data loading phase.

### 4.5.2 Reduce Frequency of Transaction Commits

A *commit* command in data loading permanently writes the loaded data to the database. The RDBMS must perform a considerable amount of processing when a transaction commits [7], but infrequent commits can lead to large redo and undo logs, and lengthen the time needed to recover the database in the event of a hardware failure. In our framework, we chose to execute commits very infrequently during the loading of catalog data, resulting in a significant performance increase.

### 4.5.3 Reduce I/O Contention

Bulk data loading is typically I/O bound. To reduce I/O contention, we distributed the database (1) data files and temporary files, (2) indices, and (3) logs onto three separate RAID devices.

### 4.5.4 Presort Data

In our framework, the data files are sorted by the primary keys of the tables prior to data loading. This sorting is done as a byproduct of the processing that extracts the catalog data from raw images. Through improvement of the clustering factor on the disk data, this presorting procedure reduces disk I/O contention, especially if Index Organized Table (in Oracle) or Clustered Index (in MS SQLServer) features are used [7].

### 4.5.5 Manage Memory Allocation

In our experiments, we discovered that allocating a smaller database data cache actually improves the data-loading performance. Since a database writer needs to scan the entire data cache when writing new data from data cache to disk, the reduced data cache size minimizes the work that the database writer has to do each time [7]. This reduced cache configuration should be adjusted after the intensive data-loading phase is complete because a larger data cache usually performs better for user queries.

## 5. Performance Analyses

In this section we report and analyze various aspects of our performance studies on parallel bulk loading of data into the Palomar-Quest sky survey repository.

All experiments were conducted in a client-server environment. All data-loading programs (clients) were implemented using Java™ and communicate with the database server using JDBC. The loading programs were initiated from multiple nodes of an NCSA Condor cluster. All nodes in the cluster are dual CPU (1.5 GHz Intel Pentium III processor), 1 gigabyte RAM, Linux servers.

An Oracle10g database runs on an 8-processor (1.3 GHz Intel Itanium 2), 64-bit SGI Altix machine running SGI Linux Propack 3 with 12 gigabytes of memory. The Altix has a single Gigabit Ethernet interface and two Qlogic FibreChannel host-bus-adapter cards. The Qlogic cards are used to access the disk environment via a storage area network (SAN). The disk environment is comprised of three separate Data Direct 8500 disk controllers, each hosting 10 terabytes of RAIDed SATA disks for a total of 30 terabytes of storage. 2-gigabit FibreChannel components are used throughout the environment.

All tests were performed using the same data model and load identical sky survey catalog data extracted from a single night's observation. All constraints, including primary key constraints, foreign key constraints, unique constraints, and check constraints were maintained in the data loading process. Tests were performed on an empty database unless otherwise noted.

The time reported is the runtime of the data-loading process as measured on the database server.

## 5.1 Bulk Loading vs. Non-Bulk Loading

Figure 4 shows the runtime of bulk loading versus non-bulk loading when a single loading process is used. For the bulk-loading process, the *bulk_loading* algorithm described earlier was used with a *batch-size* of 40. For the non-bulk loading case, a series of individual SQL insert statements were issued. The runtime of both approaches is proportional to the input data size. As shown in Figure 4, the *bulk_loading* algorithm is much faster than the individual inserts. However, although the *batch-size* was 40 we see a speedup of only 7 to 9 (not 40). This discrepancy indicates that bulk loading incurs some overhead. The *batch-size* of 40 was determined to be optimal through a series of performance tests using different values for *batch-size*.

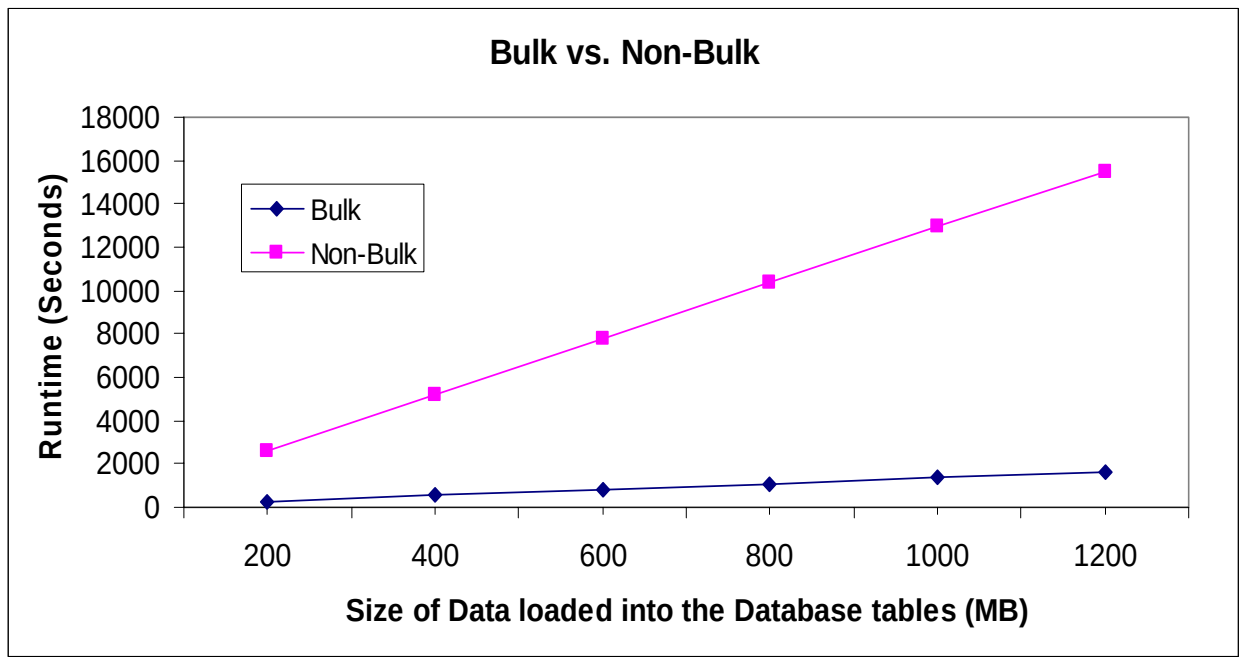

**Figure 4. Runtime of Bulk and Non-Bulk Loading**

## 5.2 Batch Size

Figure 5 shows the runtime of the *bulk_loading* algorithm with respect to the batch size used for the bulk loads. Initially, increasing the batch size decreases the loading time. However, the benefit lessens as the batch size continues to increase. The optimal batch size for the testing data set lies in the range between 40 and 50. The optimal batch size varies depending on the patterns of the catalog data file and the underlying data model. Even with this variation, experimenting with a variety of batch sizes and choosing one that is close to optimal for a typical data file can improve performance markedly over a random choice.

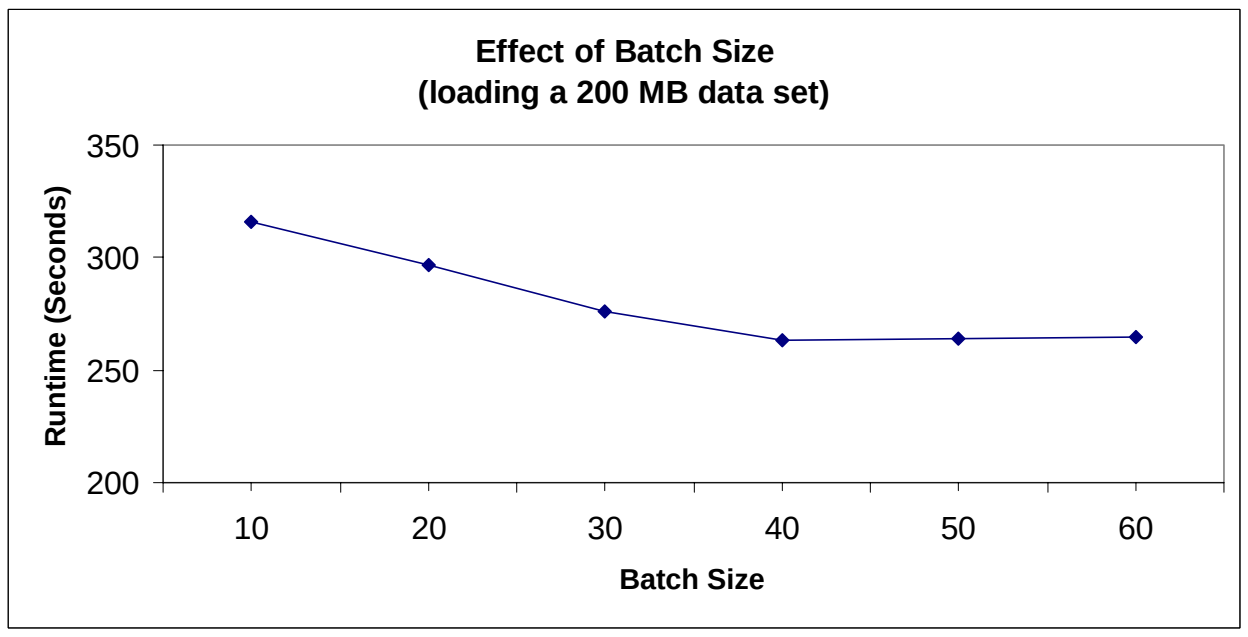

**Figure 5. Effect of Batch Size on Runtime**

## 5.3 Array Size

Figure 6 shows the runtime of the *bulk_loading* algorithm with respect to the array size. The *array-set* resides in the memory of the client host. A change to the *array-size* parameter effects the memory hit ratio and the paging rate on the client host, resulting in runtime variations on the database server. Increasing the size of the array allows more data to reside in memory and can speed up the data loading. However, on our system configuration, this benefit was lost when *array-size* was increased beyond 1000. This loss occurred because the high paging rate at those settings deteriorated the data-loading performance. These findings indicate that the *array-size* should be adjusted based on the client system configuration, the characteristics of the data model, and the interleave factor of the data being loaded. The latter two combine to determine the incremental amount of memory used when the array size is increased.

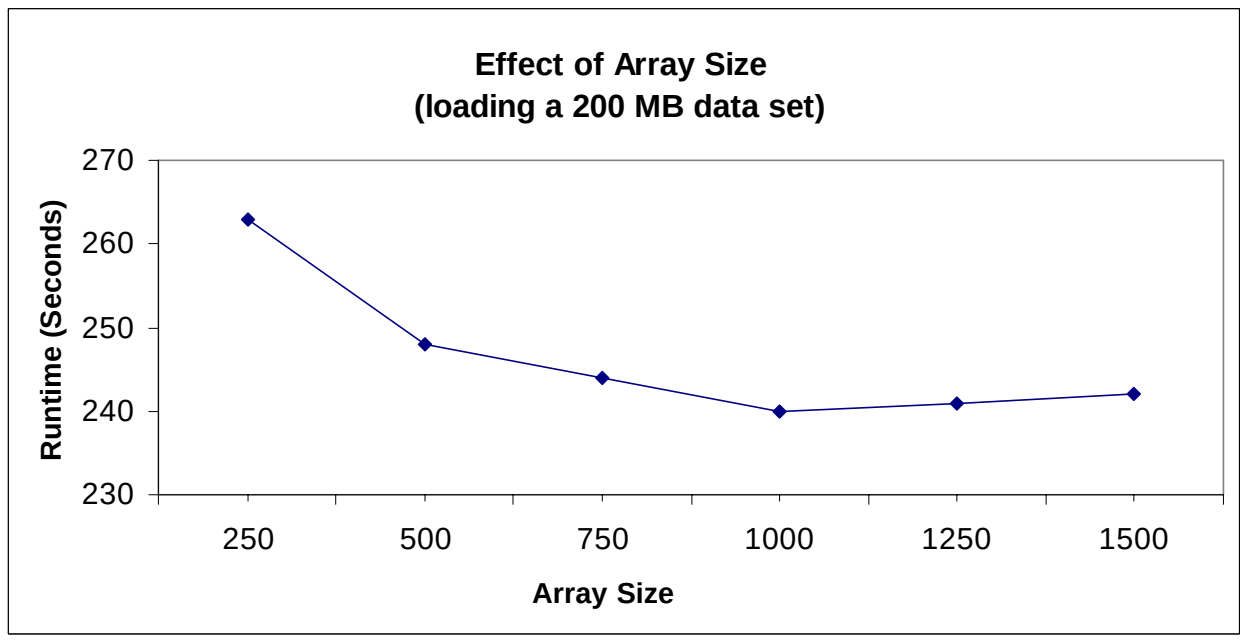

**Figure 6. Effect of Array Size on Runtime**

## 5.4 Parallel Data Loading

Figure 7 shows the loading throughput for varying numbers of concurrent bulk-loading processes running on separate nodes of the Condor cluster. A *batch-size* of 40 was used for these bulk loads. The throughput goes up almost linearly when six or fewer degrees of parallelism are used. Since the database server has eight processors, one might expect the data-loading performance should be optimal when eight loading processes are run concurrently. Unfortunately, our performance tests have shown this is not the case. The data-loading throughput peaked at a parallel degree of 6–7 in our studies. Benefits decreased after that, and, very infrequently even 6 parallel loads caused stalls and dramatic degradation in the overall throughput.

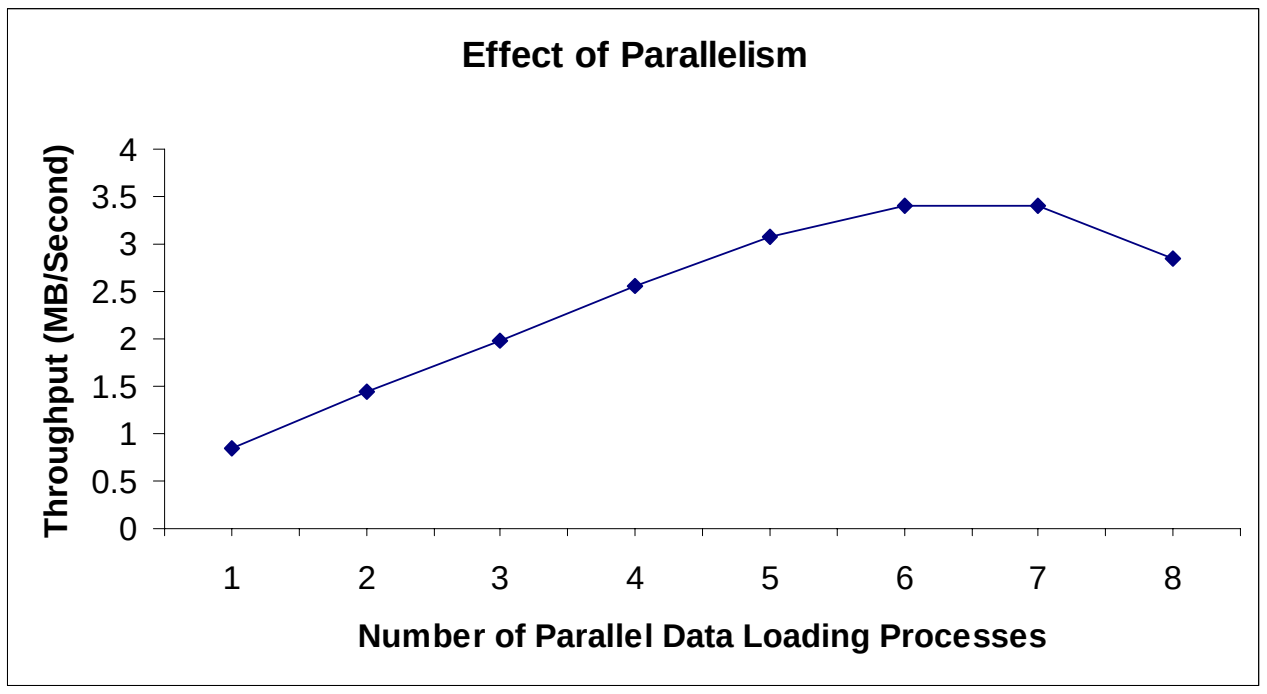

**Figure 7. Effect of Parallelism on Runtime**

As the degree of parallelism increased, we observed escalating occurrences of database locks, indicating we were hitting the RDBMS limit on the number of concurrent transactions. The relationship between the observed database locks and the database concurrent transaction limit is not an intuitive one. Our results indicate that various degrees of parallelism should be tested in the environment for typical data loading tasks to maximize the power of parallel processing while avoiding database lock contention. Because of the infrequent but very long stalls we observed when running with an "optimal" parallel degree of 6, we chose to run 5 concurrent loading processes in our production SkyLoader framework.

## 5.5 Attribute Indices

Figure 8 shows the impact of various indices on the runtime of bulk data loading. We have experimented with three scenarios. In the first, no indices were created. In the second, one index was built and maintained on a large integer attribute. In the final scenario, one index was built and maintained on three float attributes. All experiments were carried out on an empty database, with six different data set sizes bulk-loaded by a single process.

Our experiments confirm that indices do slow down the data-loading process, but to varying degrees. The single-integer attribute index had an almost undetectable average performance impact of 1.5% over the six data set sizes. In contrast, the composite index built from three float attributes causes a significant performance degradation averaging 8.5%.

In these tests, as the size of the data sets increased, the performance degradation attributable to the indices tended to increase as well. When parallel loads are being done, maintenance of indices introduces more concurrent transactions (and locks) into the bulk-load process. Furthermore, in our early prototype environment as the size of the Palomar-Quest repository grew, the overhead that the indices placed on the data-loading performance became increasingly worrisome.

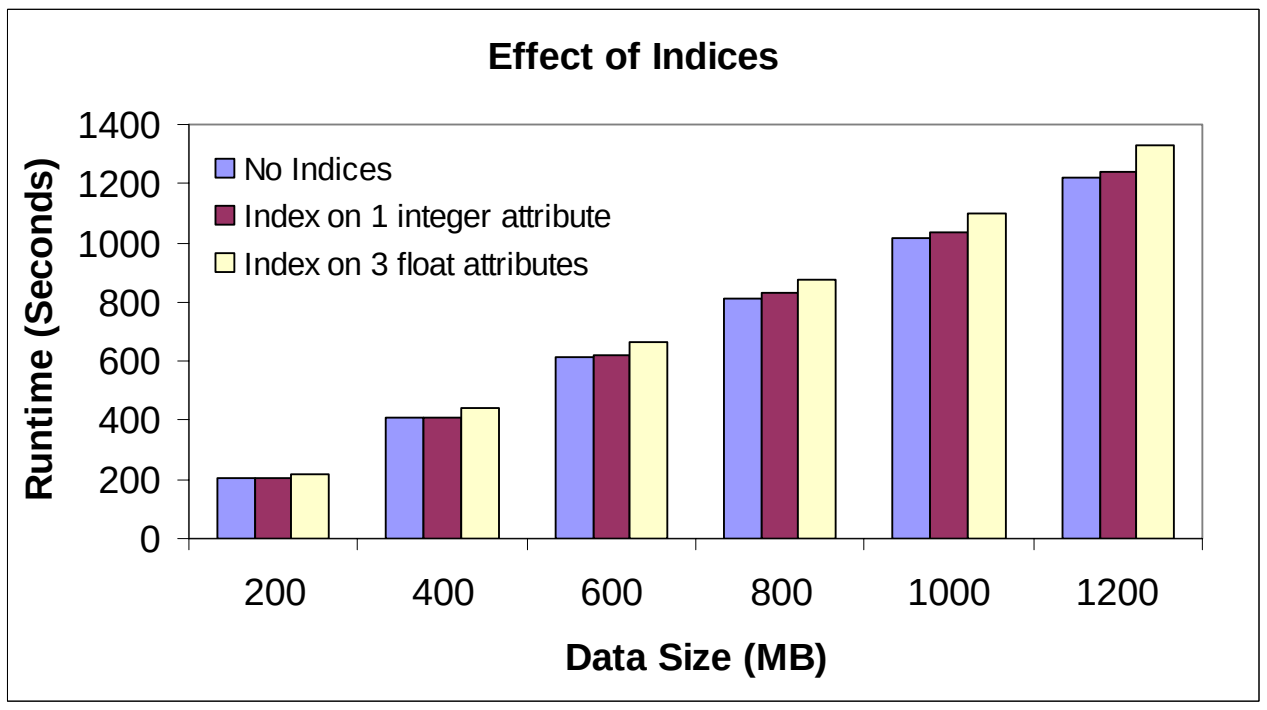

**Figure 8. Effect of Indices on Runtime**

Based on the results of these studies and our observations with the larger database, we elected to maintain only the single-integer attribute index during data loading to support user queries. With ongoing database tuning, we have been able to maintain the single-integer attribute index without incurring increased load-time overhead as the database size has grown. We have delayed the composite index creation until the intensive data-loading phase is complete.

## 5.6 Database Size

In our final performance experiment, we explored the effect of database size on bulk data-loading runtime. With some loading techniques, the time required to load the same amount of data increases as the size of the database grows. The results shown in Figure 9 indicate the database size has no significant impact on data-loading time when indices are disabled. Loading time for a 200 megabyte dataset remains constant as the size of the database grew from 50 gigabytes to 300 gigabytes. The Palomar-Quest database size currently exceeds 1.5 terabytes and we have not seen a decrease in loading speed even at this scale.

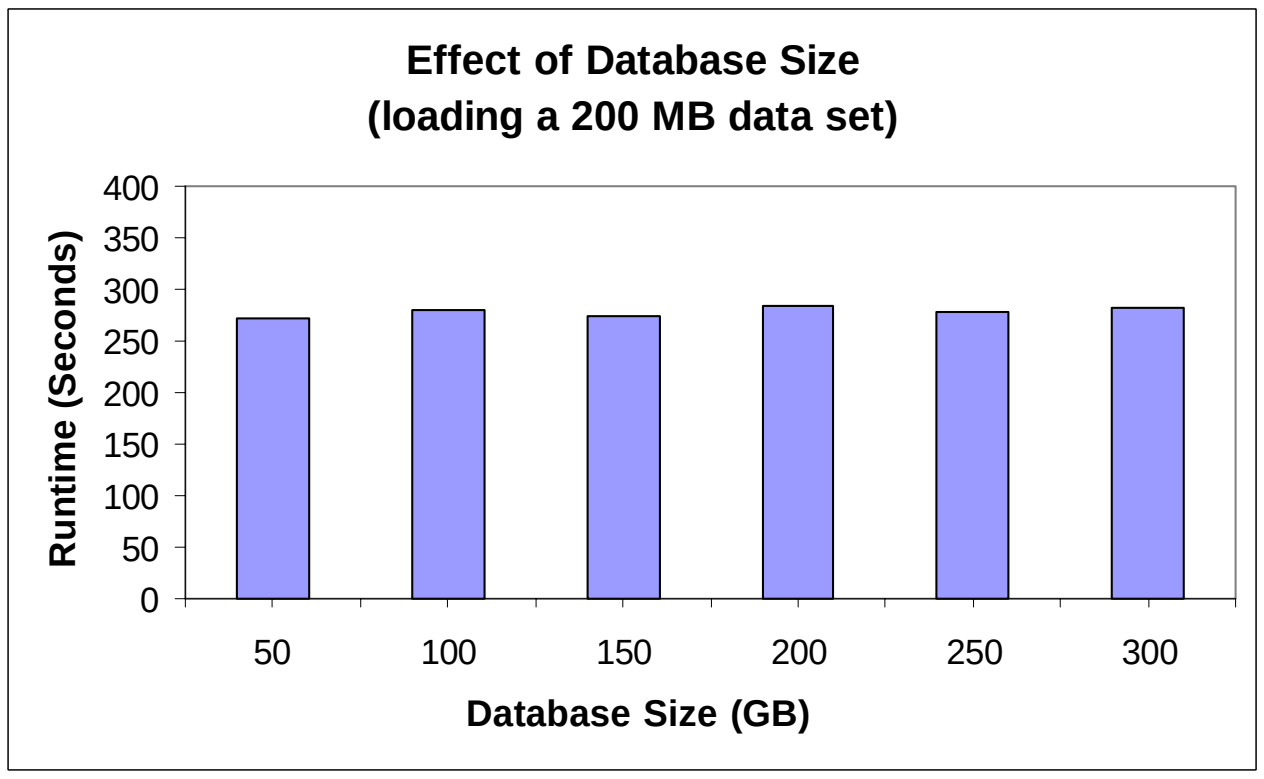

**Figure 9. Effect of Database Size on Runtime**

# 6. Discussion

Similar to other database repositories, a considerable amount of the effort that has gone into building the Palomar-Quest sky survey repository has been focused on data loading. Much time and energy has been invested in exploring new approaches to improve the data-loading performance, and in performance studies to identify optimal parameter settings for our configuration and data model. In this section we compare and contrast our data-loading approach to that of the Sloan Digital Sky Survey.

One of the most successful sky survey repositories, the Sloan Digital Sky Survey (SDSS), utilizes a framework [13, 14, 15] to perform most of the data-loading steps in parallel, on a cluster of SQL Server nodes, using distributed transactions. The SDSS data loading is a two-phase load: the data is first loaded into *Task* databases, not exceeding 20-30 gigabytes each. Then the data is fully validated before being published to its final destination in the *Publish* database. In the SDSS framework, the catalog data is converted to comma-separated-value ASCII files before the two-phase loading begins. The data in each comma-separated-value file is associated with a single database table. The data is bulk loaded into MS/SQLServer using Data Transformation Services (DTS). By carefully ordering the loading sequence of the ASCII files, the table relationships in the database are maintained.

Our framework differs significantly from the SDSS approach. In the *SkyLoader* framework, all data-loading tasks, including data validation, transformation, computation and insertion, are performed in a single pass. Since our approach does not split the data into multiple data files based on the destination database tables, nor does it require an intermediate temporary database be loaded prior to loading the permanent database, we believe our approach can be more efficient. That said, due to the incompatibility of these two repositories, we are unable to conduct a direct performance comparison to test this hypothesis.

Another difference between the SDSS data-loading process and that of the *SkyLoader* framework is that SDSS relies on a proprietary tool, DTS, while our framework is implemented using Java. The use of Java makes our framework platform-independent, portable and extensible.

## 7. Conclusions and Future Work

The Palomar-Quest sky survey is a multi-year project to collect, archive, process, and distribute survey data for research collaborations. The repository being built at NCSA to hold catalog data for the Palomar-Quest survey currently exceeds 1.5 terabytes in size, and is expected to ultimately exceed 5 terabytes. The first significant challenge this project faced was to load the catalog data into the repository database in a timely fashion. Parallel bulk loading with array buffering has proven to be a viable approach to address this challenge.

We have proposed and implemented our *SkyLoader* framework to realize this data-loading goal. Our framework consists of an efficient algorithm for bulk loading, an effective data structure to support data integrity and proper error handling during the loading process, support for optimized parallelism, and guidelines for database and system tuning. Our framework has taken advantage of high-performance computing and parallel-processing resources, and has made the building of a terabyte-plus repository a reality. With this framework, we have decreased the loading time for a 40-gigabyte data set from over 20 hours to less than 3 hours, running on the same hardware architecture and operating system.

Looking ahead, we are interested in collaborating with other scientists to apply our data-loading framework to their large scientific database problems. While some of the framework is specific to the Palomar-Quest repository, we believe that much of it is directly applicable to other domains. In particular, the knowledge gained and optimization techniques developed will allow us to quickly adapt the code to other operating environments and data models.

As part of the effort to make our framework more portable and tunable, we plan to revise the *array-set* data structure to take advantage of the memory-saving configuration options discussed in Section 4.3. The use of configuration files to control *array-set* initialization will not only lower client memory requirements, but also make the framework more adaptable for use with data sets other than the Palomar-Quest sky survey.

In addition, we plan to explore database-hosting architectures and Oracle RAC technology to see how they scale on databases of the Palomar-Quest magnitude and complexity. For many projects, the option of hosting a production database on a cluster configuration that can be scaled up as the data size and usage increases is an attractive one provided performance and stability are not sacrificed.

Finally, we will continue our collaboration on the Palomar-Quest catalog repository. With the data-loading phase under control, we will turn our attention toward tuning the database to meet the needs of the scientists who will be submitting queries through web interfaces, as well as programmatically from scientific codes executing on high-end compute resources such as those provided by the TeraGrid. Our ultimate goal is to enable new scientific discoveries through the effective coupling of compute and data technologies.

## Acknowledgements

We would like to thank Michael Remijan, Adam Rengstorf, Nicholas Waggoner, and Brian Wilhite, members of the NCSA Laboratory for Cosmological Data Mining, for their work in parallel data loading using the NCSA Condor cluster. We would also like to thank Michelle Butler, Chad Kerner, and Chris Cribbs, members of the NCSA

Storage Enabling Technologies Group, for their assistance with tuning the SGI Altix system and disk environment. Finally, we would like to thank the members of the Palomar-Quest collaboration for their dedication and hard work that has produced the rich dataset we have leveraged in the work described in this paper.

## References

[1] S. Amer-yahia and S. Cluet. "A Declarative Approach to Optimize Bulk Loading into Databases". *ACM Transactions on Database Systems*, Vol. 29, Issue 2, June 2004.

[2] T. Barclay, R. Barnes, J. Gray, P. Sundaresan. "Loading Databases Using Dataflow Parallelism". *SIGMOD RECORD*, 23(4), Dec. 1994

[3] T. Barclay, J. Gray, D. Slutz. "Microsoft TerraServer: A Spatial Data Warehouse". In Proc. SIGMOD, Austin, TX, May 2000.

[4] J. Berchen, B. Seeger. "An Evaluation of Generic Bulk Loading Techniques". In Proc. 27th VLDB Conference, Rome, Italy, 2001.

[5] C. Bohm and H. Kriegel. "Efficient Bulk Loading of Large High-Dimensional Indexes". In Proc. Int. Conf. Data Warehousing and Knowledge Discovery (DaWak), 1999.

[6] D. Burleson. "Hypercharge Oracle data load speed". http://www.dba-oracle.com/oracle_tips_load_speed.htm

[7] D. Burleson. "Hypercharging Oracle Data Loading". http://www.orafaq.com/articles/archives/000020.htm

[8] M. Graham, R. Williams, S. Djorgovski, A. Mahabal, C. Baltay, D. Rabinowitz, A. Bauer, J. Snyder, N. Morgan, P. Andrews, A. Szalay, R. Brunner, J. Musser. "Palomar-QUEST: A case study in designing sky surveys in the VO era". Astronomical Data Analysis Software and Systems XIII, ASP Conference Series. Vol. 314, 2004.

[9] S. Leutenegger, D. Nicol. "Efficient Bulk-Loading of Gridfiles". *IEEE Transactions on Knowledge and Data Engineering*, 9(3):410-420, 1997.

[10] W. O'Mullane, A.J. Banday, K.M. Gorski, P. Kuntz, A.S.Szalay. "Splitting the Sky – HTM and HEALPix", *Mining the Sky*, Banday et al ed. Springer, 2000, p639-648.

[11] A. Papadopoulos, Y. Manolopoulos. "Parallel bulk-loading of spatial data". In *Parallel Computing* 29(10):1419-1444, Oct. 2003.

[12] G. Reese. *Database Programming with JDBC and Java*. O'Reilly. 2nd Edition, Aug. 2000.

[13] A. Szalay, P. Kunszt, A. Thakar, J. Gray, R. Brunner. "Designing and Mining Multi-Terabyte Astronomy Archives: The Sloan Digital Sky Survey". In Proc. SIGMOD, Austin, TX, May 2000.

[14] A. Szalay, J. Gray, A. Thakar, P. Kunszt, T. Malik, J. Raddick, C. Stoughton, J. vandenBerg. "The SDSS SkyServer-Public Access to the Sloan Digital Sky Server Data". *Microsoft Technical Report*. MSR-TR-2001-104, Nov 2001.

[15] A. Szalay, J. Gray, A. Thakar, B. Boroski, R. Gal, N. Li, P. Kunszt, T. Malik, W. O'Mullane, M. Nieto-Santisteban, J. Raddick, C. Stoughton, J. vandenBerg. "The SDSS DR1 SkyServer, Public Access to a Terabyte of Astronomical Data". http://cas.sdss.org/dr3/en/skyserver.

[16] D. Thain, T. Tannenbaum, M. Livny. "Condor and the Grid". in *Grid Computing: Making The Global Infrastructure a Reality*. Fran Berman, Anthony J.G. Hey, Geoffrey Fox, editors. John Wiley, 2003.

[17] J. Wiener, J. Naughton. "Bulk Loading into an OODB: A Performance Study". In Proc. 20th VLDB Conference. Santiago, Chile, pp. 120–131, 1994.